\documentclass[12pt]{article}
\usepackage{feynmp,cite}
\usepackage{psfig,epsf}
\def\href#1#2{#2}

\begin{document}
\begin{fmffile}{tzc}\unitlength=1mm
\thispagestyle{empty}
 \begin{flushleft}
DESY 98--163
\\
hep-ph/9810522
\\
October 1998
 \end{flushleft}

\noindent
\vspace*{0.50cm}
\begin{center}
\vspace*{2.cm} 

{\huge 
Seeking ${\cal CP}$ Violating Couplings
\\
in $ZZ$ Production at LEP2\\
}

\vspace*{2.cm}

\renewcommand{\thefootnote}{\fnsymbol{footnote}}
{\large 
Jochen Biebel\footnote{Email: biebel@ifh.de}
}
\renewcommand{\thefootnote}{\arabic{footnote}}
\setcounter{footnote}{0}
\vspace*{0.5cm}
 
\begin{normalsize}
{\it
DESY Zeuthen,
\\
Platanenallee 6, D-15738 Zeuthen, Germany
}
\end{normalsize}
\end{center}
 
 \vspace*{2.5cm} 

\begin{abstract}

The effects of ${\cal CP}$ violating anomalous $ZZZ$ and $\gamma ZZ$ 
vertices in $ZZ$ production are determined.
We present the differential cross-section for $e^+e^-\to ZZ$
with dependence on the spins of the $Z$ bosons. 
It is shown that from the different spin combinations those with
one longitudinally and one
transversally polarized $Z$ in the final state are the most sensitive to
${\cal CP}$ violating anomalous couplings.
\end{abstract}

\vfill
\clearpage

\section{Introduction}

LEP2 runs now at energies above the $ZZ$ production threshold and the
first $Z$ pairs are observed.
During the run of LEP2 some several hundred events will be collected.
This might be enough to provide limits for $ZZZ$ and $\gamma ZZ$
vertices, which
are absent in the Standard Model of electroweak interactions
at tree-level.
However, physics
beyond the Standard Model might give strong contributions to 
neutral gauge boson vertices \cite{Zvertex}.

\begin{figure}
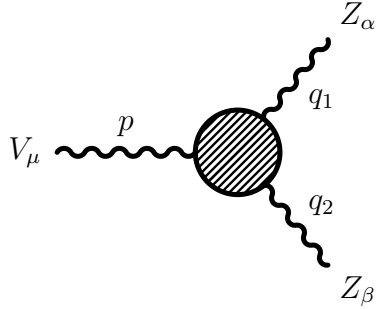

  \begin{center}
  \begin{fmfchar*}(40,30)
    \fmfpen{thick}
    \fmfleft{k1}
    \fmfright{p2,p1}
    \fmf{photon,label=$p$,l.s=left}{k1,v1}
    \fmf{photon,label=$q_1$}{v1,p1}
    \fmf{photon,label=$q_2$}{v1,p2}
    \fmflabel{$Z_\alpha$}{p1}
    \fmflabel{$Z_\beta$}{p2}
    \fmflabel{$V_\mu$}{k1}
    \fmfv{decor.shape=circle,decor.filled=.5,decor.size=16thick}{v1}
  \end{fmfchar*}
  \end{center}
  \caption{\label{tripleZ}\it The $V ZZ$ vertex with $V=\gamma,Z$.
  Two $Z$ bosons with momenta $q_i$ are on-shell,
  $q_i^2=m_Z^2$.}
\end{figure}
Requiring only invariance under Lorentz transformations
the most general $VZZ$ vertex, see figure
\ref{tripleZ}, with two
on-shell $Z$ bosons is given by \cite{Gaemers:1979hg,Hagiwara:1987vm}:
\begin{equation}
\Gamma^{\alpha\beta\mu}_{VZZ}=\frac{p^2-m_V^2}{m_Z^2}\left[
   i{f_4^V}(p^\alpha g^{\mu\beta}+p^\beta
   g^{\mu\alpha})+i{f_5^V}\epsilon^{\mu\alpha\beta\rho}
   (q_1-q_2)_\rho\right].
   \label{fvertex}
\end{equation}
In contrast to the $VW^+W^-$ vertex, with seven possible anomalous
couplings \cite{Gaemers:1979hg,Hagiwara:1987vm}, 
Bose symmetry forbids more couplings in~(\ref{fvertex})%
\footnote{%
More anomalous couplings are allowed in (\ref{fvertex}),
if one additional $Z$ boson is
off-shell, but their contributions are suppressed by a factor
$(q_1^2-q_2^2)$. 
}%
.
The anomalous parameter $f_5^V$ in (\ref{fvertex}) leads to violation
of ${\cal C}$
and ${\cal P}$ symmetry, but maintains invariance under ${\cal CP}$
transformations, while $f_4^V$ would introduce ${\cal CP}$ violation in
(\ref{fvertex}). 

The differential Standard Model cross-section for $ZZ$ production in $e^+e^-$  
annihilation is known for almost 20 years \cite{Brown:1979mq,Gaemers:1979hg}.
Also, the effects of anomalous neutral gauge boson
couplings in the production of $\gamma Z$, $ZZ$, and $\gamma$ bosons
have been studied, see e.g.~\cite{Zcalc}. 
In $Z\gamma$ production processes at the Tevatron
\cite{Tevatron
} and LEP
\cite{LEP2
}, limits to the vertex $Z\gamma Z$ were obtained.
However, they cannot be transferred to $Z$ pair production, since
the anomalous couplings in equation~(\ref{fvertex}) are independent
of couplings in $Z\gamma$ production.

\section{The Differential Cross-Section}
\label{spinamplitudes}

The differential cross-section for $ZZ$ production is presented in
this section.
It is obtained for the various spin combinations of final state $Z$ bosons.
This allows to determine their sensitivity to the anomalous couplings.

The finite width effects of the produced $Z$ bosons are considered
by convoluting the cross-section for $ZZ$ production with
two Breit-Wigner functions.
The differential cross-section is then
expressed by the sum:
\begin{equation}
 \frac{d\sigma}{d\cos\theta}=\sum_{\sigma\sigma'}
 \frac{d\sigma^{\sigma\sigma'}}{d\cos\theta},
\label{sigmasummed}
\end{equation}
with
\begin{equation}
 \frac{d\sigma^{\sigma\sigma'}}{d\cos\theta}=\int\!\mbox{d}s_1 \int\!\mbox{d}s_2\,
 \frac{\sqrt{\lambda}}{64\pi
 s^2}\left|{\cal M}^{\sigma\sigma'}\right|^2\rho(s_1)\rho(s_2),\label{sigmadiff}
\end{equation}
where $\sigma$ and $\sigma'$ stand for the possible $Z$ boson helicities
$+$, $-$, and $0$, which are defined by the vectors:
\begin{eqnarray}
\epsilon_{1\pm}&=&\epsilon_{2\mp}
=\frac{1}{\sqrt{2}}\left(0,\mp1,-i,0\right),
\label{spin1}\\
\epsilon_{10}&=&\frac{1}{2\sqrt{ss_1}}\left(\sqrt{\lambda}
 ,0,0,\left(s+s_1-s_2\right)\right),\label{spin2}
\\
\epsilon_{20}&=&\frac{1}{2\sqrt{ss_2}}\left(\sqrt{\lambda}
 ,0,0,-\left(s-s_1+s_2\right)\right)\label{spin3}.
\end{eqnarray}
The K\"all\'{e}n function $\lambda$, the Breit-Wigner factors
$\rho(s_i)$ and other notations may be inferred from \cite{Biebel:1998ww}.
The squared matrix element $\left|{\cal M}^{\sigma\sigma'}\right|^2$
in~(\ref{sigmadiff}) can be expressed by the Standard Model part
$\left|{\cal M}^{\sigma\sigma'}_{\mbox{\scriptsize SM}}\right|^2$ and
the anomalous contributions
$A^{\sigma\sigma'}$:
\begin{equation}
\left|{\cal M}^{\sigma\sigma'}\right|^2=
\left|{\cal M}^{\sigma\sigma'}_{\mbox{\scriptsize SM}}\right|^2
+A^{\sigma\sigma'}.
\label{Msplit}
\end{equation}

The squared amplitude in the Standard Model is:
\begin{equation}
   \left|{\cal M}_{\rm \mbox{\scriptsize SM}}^{\sigma\sigma'}\right|^2=
   \left(L_Z^4+R_Z^4\right)
   {\cal G}^{\sigma\sigma'}_{{\rm \mbox{\scriptsize SM}}}(s_1,s_2),
\label{MSM}
\end{equation}
where the $Zee$ couplings $L_Z$ and $R_Z$ can again be taken from
\cite{Biebel:1998ww}.
The functions ${\cal G}^{\sigma\sigma'}_{{\rm \mbox{\scriptsize SM}}}$ are:
\begin{eqnarray}
{\cal G}^{00}_{{\rm \mbox{\scriptsize SM}}}(s_1,s_2)&=&
  \frac{16s^2s_1s_2}{u^2t^2}\cos^2\theta\sin^2\theta,
\\
{\cal G}^{\pm\pm}_{{\rm \mbox{\scriptsize SM}}}(s_1,s_2)&=&
  \frac{{\cal G}^{00}}{4}+2(s_1-s_2)^2\sin^2\theta\left(\frac{1}{u^2}+
   \frac{1}{t^2}\right),
\\
{\cal G}^{\pm\mp}_{{\rm \mbox{\scriptsize SM}}}(s_1,s_2)&=&
  \frac{s^2(u+t)^2}{u^2t^2}(1-\cos^4\theta),
\\
{\cal G}^{0\pm}_{{\rm \mbox{\scriptsize SM}}}(s_1,s_2)&=&
   \frac{2ss_1}{u^2t^2}\left\{8s_2^2\cos^2\theta+4s_2\sin^2\theta\left(
   s_1-s\cos^2\theta\right)+\lambda\sin^4\theta\right\},
\\
{\cal G}^{\pm0}_{{\rm \mbox{\scriptsize SM}}}(s_1,s_2)&=&
   {\cal G}^{0-}_{{\rm \mbox{\scriptsize SM}}}(s_2,s_1),
\label{SMp0}
\end{eqnarray}
with the Mandelstam variables:
\begin{eqnarray}
   t&=&-\frac{1}{2}\left(s-s_1-s_2-\sqrt{\lambda}\cos\theta\right),\\
   u&=&-\frac{1}{2}\left(s-s_1-s_2+\sqrt{\lambda}\cos\theta\right).
\end{eqnarray}   
It is noteworthy that the coupling constant combination
$(L_Z^4-R_Z^4)$ gives additional contributions to  equation~ (\ref{MSM}).
However, these terms cancel each other for all measurable cross-sections and are
therefore not reproduced\footnote{
Since the produced $Z$ bosons are identical, not all of the various spin
combinations
are observable by themselves.
As an example, the contribution of ${\cal G}^{+-}(s_1,s_2)$ cannot
be distinguished from ${\cal G}^{-+}(s_2,s_1)$ and only their sum can
be measured.}.

The contributions from the anomalous diagrams are:
\begin{eqnarray}
   A^{\sigma\sigma'}&=&
   \sum_{V_k,V_l=\gamma,Z} \frac{1}{m_Z^4} \left(L_{V_k}L_{V_l}
   +R_{V_k}R_{V_l}\right)
   {\cal G}^{\sigma\sigma'}_{s}(s_1,s_2;V_k,V_l)
   \nonumber\\
   &&\mbox{}
   +\sum_{V=\gamma,Z}\frac{f_5^V}{m_Z^2}
   \left(L_{V}L_Z^2 -R_{V}R_Z^2\right)
   {\cal G}^{\sigma\sigma'}_{i}(s_1,s_2),
\label{MSS}
\end{eqnarray}
with the coupling constants $L_\gamma=R_\gamma=-e/2$ for the $\gamma ee$
coupling.
The denominators of the $s$-channel propagators $1/(s-m_V^2)$ cancel
with the corresponding factors
in the anomalous vertex function (\ref{fvertex}).
Photon and $Z$ exchange have the same $s$-dependence in (\ref{MSS}).
Again, as in equation~(\ref{MSM}) terms with additional coupling constant
combinations appear for the various spin combinations, but they do not
contribute to physical observables. 

The functions ${\cal G}^{\sigma\sigma'}_s$ for the $s$-channel diagrams squared are:
\begin{eqnarray}
{\cal G}^{00}_{s}(s_1,s_2)&=&
   \frac{\lambda\sin^2\theta}{4s_1s_2}\left(s_1-s_2\right)^2f_4^{V_k}f_4^{V_l},
\label{00s+}
\\
{\cal G}^{0\pm}_{s}(s_1,s_2)&=&
\frac{\left(1+\cos^2\theta\right)s}{8s_1}\left\{\lambda
f_4^{V_k}f_4^{V_l}
+(\lambda-4s_1[s-2(s_1+s_2)])f_5^{V_k}f_5^{V_l}\right\},
\\
{\cal G}^{\pm\pm}_{s}(s_1,s_2)&=&
   \left(s_1-s_2\right)^2\sin^2\theta f_5^{V_k}f_5^{V_l}, 
\\
{\cal G}^{\pm0}_{s}(s_1,s_2)&=&
   {\cal G}^{0-}_{s}(s_2,s_1),
\\
{\cal G}^{\pm\mp}_{s}(s_1,s_2)&=&0.
\label{00s-}
\end{eqnarray}
Similarly, the interferences ${\cal G}^{\sigma\sigma'}_i$ between $t$ and $u$-channel
diagram and the
$s$-channel diagram can be expressed by:
\begin{eqnarray}
{\cal G}^{0\pm}_{i}(s_1,s_2)&=&\frac{s(3s_1+s_2-s)}{ut}\left\{4s_2-
  (3s_2-s_1+s)\sin^2\theta\right\},
\label{0pi+}
\\
{\cal G}^{\pm\pm}_{i}(s_1,s_2)&=&
\frac{2(s_1-s_2)^2(u+t)}{ut}\sin^2\theta,
\\
{\cal G}^{\pm0}_{i}(s_1,s_2)&=&
    {\cal G}^{0-}_{i}(s_2,s_1),\label{pi+0}
\\
{\cal G}^{00}_{i}(s_1,s_2)&=&
{\cal G}^{\pm\mp}_{s}(s_1,s_2)=0.
\label{00i-}
\end{eqnarray}

As expected there are no contributions of anomalous couplings to the
spin combinations $(+-)$ and $(-+)$.
These are spin 2 states and cannot be produced by the $s$-channel diagrams.

Equations~(\ref{00s+}) -- (\ref{00i-}) exhibit that 
the ${\cal CP}$ violating couplings proportional to $f_4^V$%
do neither interfere with the Standard Model terms nor with terms
proportional to $f_5^V$.
A further conclusion is that it is impossible to separate out the effects
of the parameters $f_4^Z$ and $f_4^\gamma$ on the differential cross-section
of the process $e^+e^-\to ZZ$.
However, both parameters imply ${\cal CP}$ violation in a $VZZ$
vertex.

In the limit of on-shell $Z$ pair production only the combination with
one longitudinally and one transversally polarized $Z$ receives contributions
from anomalous couplings.

With the given expressions for the various spin combinations, the
differential cross-section (\ref{sigmasummed}) is obtained:
\begin{equation}
\frac{d\sigma}{d\cos\theta}=\int\!\mbox{d}s_1 \int\!\mbox{d}s_2\,
 \frac{\sqrt{\lambda}}{64\pi s^2}\rho(s_1)\rho(s_2)\left\{S+A_i+A_s\right\},
\label{comp}
\end{equation}
with the functions
\begin{eqnarray}
S\!&=&\!
2\frac{L_Z^4+R_Z^4}{u^2t^2}\left\{
    4\lambda \sin^2\theta ( s\sigma + s_1s_2 )
    + \lambda^2(1-\cos^4\theta) +  16ss_1s_2\sigma  \right\},
\label{Sdef}\\
A_i\!&=&\!-\frac{4}{m_Z^2}\sum_{V=\gamma,Z}\!\!\frac{L_{V}L_Z^2
-R_{V}R_Z^2}{ut}f_5^V\left\{
 2s[(s-\sigma)\sigma-4s_1s_2 ]
-\lambda(s+\sigma)\sin^2\theta\right\},
\\
A_s\!&=&\!
\sum_{V_k,V_l=
\gamma,Z}\!\!\frac{L_{V_k}L_{V_l}+R_{V_k}R_{V_l}}{4s_1s_2m_Z^4}\left\{
   \lambda (1+\cos^2\theta)
    \left[   s\sigma f_4^{V_k}f_4^{V_l} +(s\sigma-8s_1s_2) f_5^{V_k}f_5^{V_l}
     \right]   
\label{ASdef}
   \right.\\&& \mbox{}\left.
\hphantom{\frac{L_{V_k}L_{V_l}+R_{V_k}R_{V_l}}{4s_1s_2m_Z^4}}   
     +(s_1-s_2)^2\left[  \lambda\sin^2\theta f_4^{V_k}f_4^{V_l}
       + 16s_1s_2f_5^{V_k}f_5^{V_l}\right] \right\},\nonumber
\end{eqnarray}
where the abbreviation $\sigma=s_1+s_2$ is used.
The Standard Model part is described by $S$ while the anomalous
contributions are contained in $A_i$ and $A_s$. 

In the Standard Model limit the differential cross-section in
(\ref{comp}) is in agreement with the result presented 
in \cite{Leike:1995kj}.
Radiative corrections to (\ref{comp}) due to initial state radiation, see e.g.
\cite{Bardin:1996jw},
can be applied in the structure function approach including a Lorentz boost
of the scattering angle as described similarly in $W$ pair production
\cite{Biebel:1998ww}.

\section{Conclusions}

The presented analytical results demonstrate how potential anomalous
couplings change the differential $ZZ$ production cross-section.
Anomalous couplings will have their main effect in the
production of a longitudinally and a transversally polarized $Z$ boson.
For on-shell $ZZ$ production, these are the only spin combinations
which are sensitive to an anomalous signal.
Effects of non-vanishing $f_4^V$ and $f_5^V$ on the
final states with two transversally or two longitudinally polarized
$Z$ bosons are zero or are suppressed by a factor $(s_1-s_2)^2$.

Therefore, a measurement of the final state spins might be used to
increase the ratio of the anomalous signal over the Standard Model 
background.
But, since only one type of spin combinations is sensitive to
anomalous couplings, 
a spin analysis cannot be used to disentangle the signals from
$f_4^V$ and $f_5^V$. 
However, due to the interference contribution of $f_5^V$  in (\ref{0pi+})
and (\ref{pi+0}),
the anomalous parameters $f_4^V$ and $f_5^V$ show different angular
distributions and it might be possible to
separate out their effects at LEP2.

\bigskip

\noindent
{\large\bf Acknowledgment}
\nopagebreak\\
I would like to thank T.~Riemann for many useful discussions and
suggestions.
In addition, I am grateful to T.~Hebbeker and P.~Moln\'{a}r for
drawing my attention to the problem, to J.~Alcaraz for numerical
comparisons and to W.~Lohmann
for providing me with information about the current experimental situation.
Finally, I am also grateful to J.~Illana and T.~Riemann for carefully
reading the manuscript.

\end{fmffile}
\end{document}